\documentclass[fleqn, usenatbib]{mnras}
\usepackage[T1]{fontenc}
\usepackage{ae,aecompl}
\usepackage{graphicx}
\usepackage{color}
\usepackage{amsmath}

\title[Schwarzschild models of mock dwarf galaxies]
{Recovering the mass profile and orbit anisotropy of mock dwarf galaxies with Schwarzschild modelling}

\author[K. Kowalczyk et al.]{
Klaudia~Kowalczyk$^{1}$\thanks{E-mail: \href{mailto:klaudia.kowalczyk@gmail.com}{klaudia.kowalczyk@gmail.com}},
Ewa L.~{\L}okas$^{1}$ 
and Monica~Valluri$^{2}$ 
\\
$^{1}$Nicolaus Copernicus Astronomical Center, Polish Academy of Sciences, Bartycka 18, 00--716 Warsaw, Poland\\
$^{2}$Department of Astronomy, University of Michigan, 1085 South University Ave., Ann Arbor, MI 48109, USA
}

\pubyear{2017}

\begin{document}
\label{firstpage}
\pagerange{\pageref{firstpage}--\pageref{lastpage}}
\maketitle

\begin{abstract}
We present a new study concerning the application of the Schwarzschild orbit superposition method to model spherical
galaxies. The method aims to recover the mass and the orbit anisotropy parameter profiles of the objects using
measurements of positions and line-of-sight velocities usually available for resolved stellar populations of dwarf
galaxies in the Local Group. To test the reliability of the method, we used different sets of mock data extracted from
four numerical realizations of dark matter haloes. The models shared the same density profile but differed in
anisotropy profiles, covering a wide range of possibilities, from constant to increasing and decreasing with radius.
The tests were done in two steps, first assuming that the mass profile of the dwarf is known and employing the method to
retrieve the anisotropy only, and then varying also the mass distribution. We used two kinds of data samples:
unrealistically large ones based on over 270\,000 particles from the numerical realizations and small ones matching the
amount of data available for the Fornax dwarf. For the large data samples we recover both the mass and the anisotropy
profiles with very high accuracy. For the realistically small ones we also find a reasonably good agreement between the
fitted and the input anisotropies, however the total density profiles can be significantly biased as a result of their
oversensitivity to the available data. Our results therefore provide convincing evidence in favour of the applicability
of the Schwarzschild method to break the mass-anisotropy degeneracy in dwarf galaxies.
\end{abstract}


\section{Introduction}

Dwarf galaxies are believed to be the most dark matter dominated objects in the Universe with dark to baryonic mass
ratios even of the order of hundreds (\citealt{mateo_1998}, \citealt{gilmore_2007}) so they seem to be the best
laboratory for studying this unexplored component of the Universe. For decades astronomers have been running
simulations of the behaviour of dark matter (\citealt{navarro_1995}, \citealt{springel_2005}, \citealt{diemand_2008})
in order to describe the structure formation, profiles of dark matter distribution, sizes and shapes of dark haloes and
compare the resulting observables with the astronomical data to find the best model describing dark matter.

However, we do not have at our disposal reliable tools even to precisely measure the most basic parameter of a dwarf
galaxy, its mass. The most commonly used Jeans modelling \citep{GD} based on fitting of the velocity dispersion is
subject to the mass-anisotropy degeneracy \citep{binney_1982}, the degeneracy between the underlying mass profile and
the anisotropy of orbits of the tracer particles, since the anisotropy profile is generally unknown. 
The degeneracy can be partially lifted by including kurtosis into the
fit (\citealt{lokas_2002}, \citealt{lokas_2005}, \citealt{richardson_2013}). The method however relies 
on the predefined form of the anisotropy profile. The standard assumption is then for the anisotropy to be constant with 
radius. This is much more restrictive than the range of possibilities that are found in simulations (\citealt{campbell_2017}, 
\citealt{elbadry_2017}) e.g. profiles that are monotonically growing or decreasing with radius, and interestingly, also 
profiles with a pronounced local maximum.

A powerful tool to break the mass-anisotropy degeneracy is the application of stellar proper 
motions. \citet{wilkinson_2002} and \citet{strigari_2007} showed that 100-200 measurements are sufficient in order to 
lift the degeneracy in Jeans modelling method. Unfortunately, we still do not have such results at 
our disposal as the Space Interferometry Mission (SIM), on which these authors based their studies, has been cancelled. 
Proper motions of stars in the nearby dwarf galaxies could be also obtained with MICADO instrument \citep{trippe_2010}. 
However, in this case we may expect first measurements by 2030. Since, as pointed out by \citet{majewski_2008}, Gaia 
mission does not have capabilities necessary to derive proper motions of stars in even the nearest dwarf galaxies with high enough 
precision, the prospect of breaking the mass-anisotropy degeneracy in this way will not be realized in the near future.

In contrast, the growing availability of sensitive multiobject spectrogaphs on 6-8m class telescopes is making it possible 
to greatly increase the number of radial velocity measurements in nearby dwarf spheroidal galaxies. It is therefore worthwhile 
investigating how well both anisotropy and mass profiles can be recovered with much larger radial velocity samples than 
currently available.

One method which does not require the prior knowledge of the orbit anisotropy is the orbit superposition modelling,
first introduced by \citet{schwarzschild_1979} for constructing distribution functions for triaxial galaxies. It was
first applied to modeling kinematical data in spherical galaxies by \citet{richstone_1984} and has been developed since
then in the studies of massive early type galaxies and bulges of spiral galaxies, in order to derive their mass
distribution and mass-to-light ratios and/or to infer the existence of the black holes in the centres of ellipticals
and measure their masses (\citealt{vdMarel_1998}, \citealt{cretton_1999}, \citealt{gebhardt_2003},
\citealt{valluri_2004}, \citealt{thomas_2004}, \citealt{cretton_2004}, \citealt{cappellari_2006},
\citealt{vdBosch_2010}). The first proposed models described simplest, spherical objects (\citealt{richstone_1984},
\citealt{rix_1997}) but complexity of studies increased with time (as better data and higher computational power
were emerging) going through 3-integral axisymmetric models \citep{vdMarel_1998} up to triaxial \citep{vdBosch_2010}.

The application of the method to dwarf galaxies has been attempted only recently. First of all, the methodology
needs to be adapted because, in contrast with luminous ellipticals, dwarfs seem to be dominated by dark matter at all scales, its
spatial distribution not necessarily following the distribution of the visible tracer (stars). The subject of the 
form of the dark
matter density profile has been a matter of extensive study over the last few decades. It is still under debate whether
these profiles should be modelled as cuspy Navarro-Frenk-White (NFW, \citealt{NFW_1997}) or Einasto \citep{ludlow_2013}
profiles emerging from cosmological, dark matter only simulations or rather by a variety of cored profiles resulting
from simulations including baryonic physics \citep{governato_2010}. Moreover, Jeans modelling of dwarf
spheroidal (dSph) galaxies tends to suggest that they do not share a universal dark matter profile
\citep{walker_2009}. Therefore the determination of not only the total mass but also its distribution in dwarf galaxies
is currently one of the hottest topics in galactic dynamics.

The Schwarzschild modelling method has been applied to dSphs of the Local Group (LG) in order to obtain
density profiles of their dark matter haloes in the case of Fornax and Draco by \citet{jardel_2012} and
\citet{jardel_2013} and independently by \citet{breddels_2013b} for Fornax, Sculptor, Carina and Sextans.
Unfortunately, the results are not conclusive. Whereas \citeauthor{breddels_2013b} find cuspy profiles to be
favoured over cores for all galaxies in their sample (except for Sextans but in this case the fit was done
for only two data bins), for Fornax \citeauthor{jardel_2012} reject the NFW profile at a high confidence level.
On the other hand, according to \citeauthor{jardel_2013} Draco is embedded in an NFW-like halo.

The differences between modelling ellipticals/bulges and dwarfs of the LG also concern the
types of data available. In the case of ellipticals/bulges we deal with the integrated light distribution
and line-of-sight velocity profiles which need to be extracted from the integrated stellar spectra. In dwarfs
we are able to resolve individual stars and measure their positions and line-of-sight velocities, which then require 
different treatment. \citet{chaname_2008} developed a  maximum-likelihood based version of the 
Schwarzschild method that allows the orbit libraries to fit these individual positions and velocities. Unfortunately, 
to date this method has not been applied extensively to real data (however see \citealt{breddels_phd}). The two main 
approaches that have been applied to dwarf galaxies resort to binning the data in radius: one relies on using velocity 
moments (\citealt{breddels_2013},
\citealt{breddels_2013b}) and the other on fitting the full line-of-sight velocity
distribution (\citealt{jardel_2012}, \citealt{jardel_2013}). Both have some disadvantages: the former necessarily
leads to the loss of some information while the latter has to struggle with large errors.

With so many conflicting results and discrepancies, the reliability of the Schwarzschild orbit superposition method
needs to be tested on mock data. Such an experiment has been performed by \citet{breddels_2013} but only for one mock
numerical realization of a Sculptor-like galaxy with an adopted anisotropy. In this work we therefore intend to
investigate the reliability of the method trying to recover the mass profile of mock dark matter haloes (as a
first approximation of a dwarf galaxy) for a larger variety of anisotropy profiles. Moreover, we examine the ability
of our orbit superposition code to adequately recover the anisotropy profile as a result
rather than an assumption of the modelling method. This will enable us to verify if the Schwarzschild modelling is
truly independent of the intrinsic anisotropy and also to determine whether it is capable of recovering the
velocity anisotropy profile, thereby breaking the mass-anisotropy degeneracy.

The paper is organized as follows. In section \ref{data} we describe the numerical models used, in section
\ref{technicals} we introduce the Schwarzschild modelling scheme, in section \ref{large_sample} we carry out the
recovery of the anisotropy and mass profiles for large data sets and in section \ref{small_sample} we apply the method
to a data sample typical for a dwarf galaxy.
We summarize our results in section \ref{summary} and discuss them
in the light of the available literature and current state of knowledge in section \ref{discussion}.

\section{Mock data}
\label{data}

\subsection{Numerical realizations}
\label{realizations}

In our study we use numerical realizations of dark matter haloes containing $10^6$
particles and generated using the distribution function of \citet{wojtak_2008}. The spherically symmetric density
profile of the haloes is given by the formula:
\begin{equation}
\rho(r) = \left\{
\begin{array}{ll}
\frac{\rho_0}{(r/r_s)(1+r/r_s)^2} & r<r_v\\
\frac{N}{(r/r_c)(1+r/r_c)^5} & r>r_v\\
\end{array} \right.
\label{eq:mass_profile}
\end{equation}
corresponding to the cuspy, cosmologically motivated NFW profile (with $\rho(r)\propto
r^{-1}$ at the centre and $\propto r^{-3}$ at infinity) within the virial radius $r_v$ and the steeper
cut-off $\rho(r) \propto r^{-6}$ beyond. The cut-off is necessary to ensure the finite mass of the halo.

We use models with the virial mass $M_v=10^9 $M$_{\sun}$ and concentration $c=20$, which translate to the
following values used in eq.~(\ref{eq:mass_profile}): $\rho_0=1.77\times10^7$\,M$_{\sun}$\,kpc$^{-3}$,
$r_v=25.80$\,kpc, $r_s=1.29$\,kpc, $N=1.36\times10^4$\,M$_{\sun}$\,kpc$^{-3}$ and $r_c=41.92$\,kpc (see
\citealt{lokas_2001} for the discussion on the dependence of the NFW profile on the cosmological model in use). 
Consequently each dark matter particle has a mass of $1533.83$\,M$_{\sun}$.

The models differ in the underlying orbit anisotropy profile, defined as:
\begin{equation}
\label{eq:beta}
 \beta(r)=1-\frac{\sigma_{\theta}^2(r)+\sigma_{\phi}^2(r)}{2\sigma_r^2(r)}
\end{equation}
where $\sigma_{r,\,\theta,\,\phi}$ are the components of the velocity dispersion in the spherical coordinate system
with the origin at the centre of the halo.

We use four models in total, two with anisotropy constant with radius, $\beta=0$ and $\beta=0.5$,
and two with varying anisotropy: growing and decreasing from 0 (0.5) at the centre of the halo to 0.5 (0) at
infinity, reaching an intermediate value $\beta=0.25$ at $r_s$.

\subsection{Tracer particles}
\label{tracer}
In the observed dwarf galaxies stars are most probably distributed differently than dark matter. Dark matter haloes are
believed to be more extended and in our models we describe them by the NFW profile, whereas the observed (projected)
stellar profiles are best fitted with more concentrated profiles like Plummer \citep{plummer_1911}, S\'ersic
\citep{sersic_1968} or King \citep{king_1962}.

In order to test our method on more realistic data, we decided to select subsamples of particles following the
S\'ersic profile:
\begin{equation}
 I(R)=I_0{\rm exp}[-(R/R_s)^{1/m}],
\end{equation}
where $I_0$ is the normalization, $R_s$ is the characteristic radius and $m$ is the S\'ersic index.

Introducing the stellar component into a dark matter-only models by taking subset of dark matter 
particles follows the practice used by other studies. \citet{bullock_2005} and 
\citet{penarrubia_2008} suggested a method of selecting such subsamples in dynamical equilibrium. However,
the method is relatively simple only for an isotropic halo. Therefore we applied another approach instead.
In each halo we selected 273\,078 particles reaching as far as $34$\,kpc given with the deprojected S\'ersic profile
\citep{lima_1999}:
\begin{equation}
\nu(r)=\nu_0\Big(\frac{r}{R_s}\Big)^{-p}{\rm exp}\bigg[-\Big(\frac{r}{R_s}\Big)^{1/m}\bigg]
\end{equation}
where:
\begin{equation}
 p=1-0.6097/m+0.05463/m^2
\end{equation}
with $R_s=0.4$\,kpc and $m=1.6$. The normalization $\nu_0$ was chosen so that the maximum number of particles
could be used. Then, we evolved the haloes in isolation using the $N$-body code GADGET-2 \citep{springel_2005}
following the selected particles, until equilibrium was reached.

In the following we will assume that the total mass of the stars is at first approximation
negligible in comparison with the mass of the dark matter halo. The selected subsamples of particles remain 
dark matter particles and no particles (of any type) have been added to 
the systems. Therefore, our simulations do not contain stellar particles and denoting the subsamples as {\it stars} is 
just a convention, since their spacial distribution and kinematics can be identified as the distribution and 
kinematics of the massless tracer. Nevertheless, the marked particles' physical mass still contributes to the total 
mass of the {\it dark matter} halo.

\begin{figure}
\includegraphics[trim=35 0 0 0, clip, width=\columnwidth]{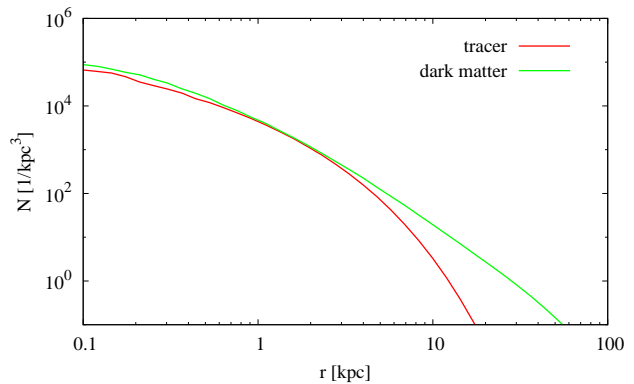}
\caption{The number density profiles of the tracer particles (red line) and all particles of the halo (green
line). The profiles overlap at around 1 kpc as a result of the choice of normalization for the tracer profile so that
the maximum number of particles could be used.}
\label{fig:profile}
\end{figure}

In Fig.\ref{fig:profile} we compare the number density profiles of the selected tracer particles in red and all
particles marked as `dark matter' in green. In the central part of a halo profiles are similar but the S\'ersic
profile of stars becomes steeper at $r\sim 1$\,kpc and drops quickly with radius.

As shown in \citet{gajda_2015} (where the models we use are labelled C1, C3, I2 and D) who evolved the haloes
in isolation for 10\,Gyr, the models with $\beta=0$ and growing anisotropy are stable and remain spherical
till the end of the simulations. When considering our subsamples only, the particles achieved equilibrium,
i.e. the moment when the profiles of density, radial velocity dispersion and anisotropy stopped changing, after
approximately 6\,Gyr but we let them evolve for another 2\,Gyr and took for further analysis the outputs at
8\,Gyr. However, the models with $\beta=0.5$ and decreasing anisotropy are subject to radial orbit instability and
become significantly non-spherical in less than 1\,Gyr. In those cases we took the outputs at 0.6\,Gyr, keeping in mind
that they are not in exact equilibrium. In all chosen outputs the total density profiles remained unchanged.

\begin{figure}
\includegraphics[trim=50 40 110 45, clip, width=\columnwidth]{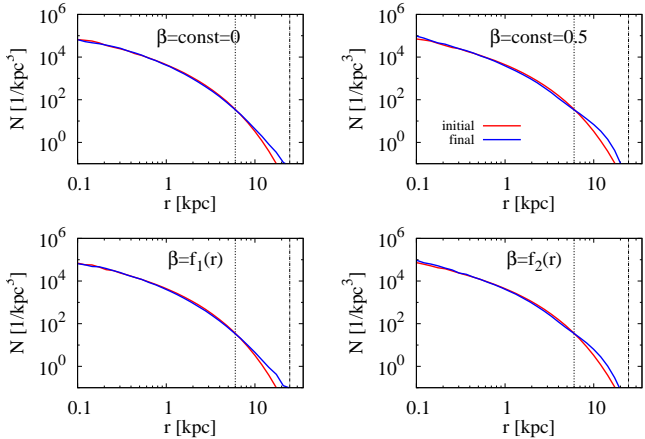}
\caption{The number density profiles of the tracer particles in the initial conditions (red lines) and final
outputs used for further analysis (blue lines) as a function of the radius from the centre. The thin dashed and
dashed-dotted vertical lines indicate the adopted upper radial limit for the mock data and the outer radius of the
orbit library, respectively.}
\label{fig:profiles_init_fin}
\end{figure}

\begin{figure}
\includegraphics[trim=50 40 110 55, clip, width=\columnwidth]{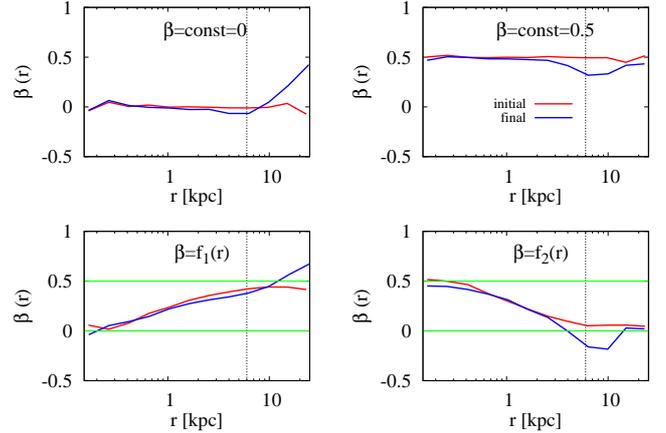}
\caption{The profiles of the anisotropy of the tracer particles in the initial conditions (red lines) and final
outputs used for further analysis (blue lines) as a function of the radius from the centre. The thin dashed vertical
lines indicate the adopted upper radial limit for the mock data. The asymptotes of the initial varying $\beta$ profiles
are marked in green.}
\label{fig:beta_init_fin}
\end{figure}

After the evolution in isolation the properties of the selected particles differ slightly from the initial values,
however the general behaviour remains the same in each case. We compare the number density profiles in
Fig.\,\ref{fig:profiles_init_fin} and the anisotropy profiles calculated with eq.\,(\ref{eq:beta}) 
in Fig.\,\ref{fig:beta_init_fin}, where the values from
the initial conditions are presented in red and final values in blue for each model separately. The thin dashed
vertical lines indicate the upper limit of the radius for the mock data projected along the line of sight which we
adopted to be $R=6$\,kpc. Additionally in Fig.\,\ref{fig:profiles_init_fin} the dashed-dotted lines indicate the
outermost radius of our orbit library (see section \ref{orbits}) and in Fig.\,\ref{fig:beta_init_fin} the green lines
mark the asymptotes of the initial profiles. The changes in the profiles of anisotropy at the outer data radii are
adventitious as the parameters of the initial S\'ersic profile were chosen after deciding on the data range (based on
the anisotropy profiles). In all of the following figures we will refer to the initially constant models as $\beta=0$
and $\beta=0.5$ regardless of the variations whereas the models with the varying anisotropy will be labelled as
$f_1(r)$ and $f_2(r)$ for the growing and decreasing profiles, respectively.

\section{The method}
\label{technicals}

In this section we present the steps of the Schwarzschild method and the parameters used for the modelling which need
to be adjusted to the data, mainly their size constraining a number of spatial bins which can be used and the spatial
distribution constraining the energy range of the orbit library.

Our approach is based on the original Fortran code by \citet{valluri_2004} developed as a complete tool to model
elliptical, axisymmetric galaxies. The code generates initial conditions for the orbit library by deriving the
multipole expansion of the potential from a given mass profile, integrates the orbits and stores their observables,
reads the observational data and fits the constraints. However, the methodology for dwarfs differs so much that we were
unable to apply the code in a straightforward way. Therefore, we have used only the first part of the original code
which generates the initial conditions and we modified (simplified) it for use in the spherical case. The software
necessary for the later stages of the modelling has been written especially for the purpose of this work in
C\texttt{++}.

\subsection{Orbit library}
\label{orbits}

For the purpose of the application of the method we generated a library of 1\,200 orbits sampling the energy and
angular momentum spaces. We used 100 values of energy in units of the radius of the circular orbit sampled
logarithmically and 12 values of the relative angular momentum $l=L/L_{\rm max}$, where $L_{\rm max}$ is the angular
momentum of the circular orbit, sampled linearly within the open interval $l\in(0, 1)$ to avoid numerical errors. The
initial conditions for the orbits were calculated under the assumption that each particle was placed at the apocentre
of its orbit. The apocentres in the library fall between $r_{\rm in}=0.081$\,kpc which is smaller than the upper limit
of the innermost bin of the constraints (see the next section) and $r_{\rm out}=24.656$\,kpc, which is $\sim4.1$ times
larger than the outer boundary of the mock observations and contains over $99.9\%$ of the tracer particles
$\Big(\frac{N(r_{out})}{N(r\rightarrow\infty)}\ge 0.999\Big)$.

The orbits have been integrated simultaneously in two groups using the $N$-body code GADGET-2, modified to accommodate
a constant potential (by adding for each particle in each timestep accelerations calculated from the Gauss theorem for
a given mass profile), for $t=10$\,Gyr (the inner 600 orbits) or $t=100$\,Gyr (the remaining ones) in order to cover at 
least a few full orbital periods even for the most extended orbits. In each case we saved 2\,001 outputs in equal timesteps. 
By definition, the library contains a set of orbits of test particles, i.e. the massless tracers of 
the underlying potential. We applied the $N$-body code instead of a standard numerical integration scheme because of its 
numerical convenience and speed. However, as the code requires positive masses of particles, we guaranteed that their 
gravitational interaction did not affect the resulting orbits by assigning a very small mass to the particles. We confirmed 
that it was sufficient and the approach did not need to be changed. Storing 
the actual orbits and not only observables demands more disk space but allowed us to reuse the orbits and recalculate 
the library with different spatial binning, saving computational time.

\subsection{Extracting observables}
\label{observables}

\begin{figure}
\includegraphics[width=\columnwidth]{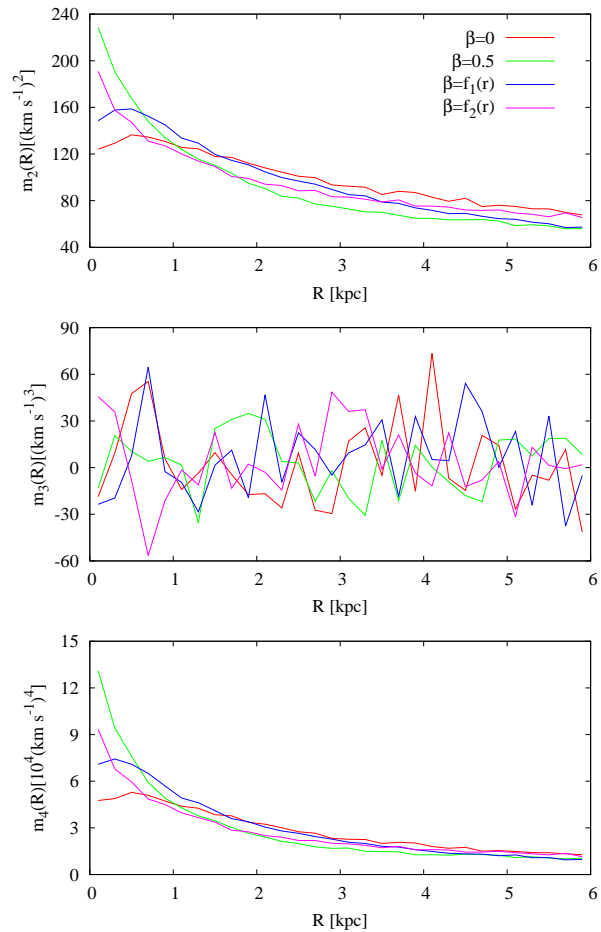}
\caption{The values of the 2nd, 3rd, and 4th velocity moment (top to bottom panels, respectively) for the four models:
$\beta=0$ (red), $\beta=0.5$ (green), increasing $\beta$ (blue) and decreasing $\beta$ (magenta) measured using all
tracer particles.}
\label{fig:mom_all}
\end{figure}

In order to generate mock data sets, we observed each galaxy, i.e. the particles marked as stars, along an arbitrarily
chosen axis, saving projected positions of the particles in terms of their distances from the centre and the
line-of-sight velocities, setting the outer maximum projected radius of the data to $R=6$\,kpc to imitate the
distribution of stars in a real dwarf galaxy. We binned the data in 30 radial bins spaced linearly. In each radial bin
we derived the proper moments of velocity: the second ($m_2$), third ($m_3$) and fourth ($m_4$), calculated with
estimators based on the sample of $N$ line-of-sight velocity measurements $v_i$ \citep{lokas_2003}:
\begin{equation}
 m_{n, l}=\frac{1}{N_l}\sum_{i=0}^{N_l}(v_{i}^{l}-\bar{v_l})^n
 \label{eq:mom}
\end{equation}
where
\begin{equation}
 \bar{v_l}=\frac{1}{N_l}\sum_{i=0}^{N_l}v_{i}^{l}
 \label{eq:mean}
\end{equation}
and $l$ labels the radial bins. We present the resulting moments in the three panels of Fig.\,\ref{fig:mom_all} with
colours denoting different anisotropy models: red for the isotropic case $\beta=0$, green for $\beta=0.5$, blue for the
growing anisotropy and magenta for the decreasing one. The noise, clearly visible for $m_3$, is the consequence of the
dense spatial binning.

The kinematics of a galaxy can be also expressed in the terms of the Gauss-Hermite moments $h_i$
(\citealt{vdMarel_1993}, \citealt{gerhard_1993}). We have tested their application on the original data i.e. unevolved
dark matter haloes, and presented the
results in \citet{kowalczyk_2016}. They proved to be useful in recovering the anisotropy profiles but at the same
time demanded large amounts of data to derive the moments correctly. As the studies
of dwarf galaxies struggle with rather limited data samples, we have abandoned this approach.

In the next step we need to obtain observables for the orbits from the library. As the orbits in the spherical
potential are coplanar we randomly rotated each orbit 100\,000 times around two axes of the simulation box and combined
them to mimic the symmetry. We have noticed that the value of the resulting $\chi^2$ (see next section) depends on this
random choice, i.e. on a particular set of rotation angles, therefore revealing a numerical defect of the method,
as a finite set of rotation angles is not sufficient to achieve the needed level of symmetry. In a set of tests we
have established that the optimal number of rotations is 100\,000 for which the $\chi^2$ varies by less than 0.1 between
the selections and the computational time necessary for rotations is reasonably short.

The orbits have been observed along an arbitrarily chosen axis and their observables have
been stored on the same grid as the mock data. The velocity moments have been calculated following eq.\,(\ref{eq:mom})
and (\ref{eq:mean}).

For the purpose of recovering the anisotropy we have also stored the three components of the velocity dispersion in
spherical coordinates as a function
of the deprojected radius in 30 linearly spaced bins in the range $r\in[0, 6]$\,kpc for the mock data and for the
orbits.

\subsection{Fitting of constraints}

The Schwarzschild method is based on the assumption that the observed kinematics of a galaxy (or in our case of marked
particles of a dark matter halo) can be reproduced as a linear combination of the same parameters for the orbits from
the library by assigning non-negative weights $\gamma$ to the orbits. Our approach is a combination 
of procedures proposed by other authors (\citealt{rix_1997}, \citealt{valluri_2004}, \citealt{vdBosch_2008}, 
\citealt{breddels_2013}) developed in order to obtain more efficient method without imposing unnecessary assumptions limiting 
its application. The fitting is performed by minimizing the deviation between the data and the
linear combination of orbits in $\chi^2$ sense via the orbital weights. The general function to minimize is:
\begin{equation}
\label{eq:fit}
 \chi^2=\sum_{l}\sum_{n}\Bigg(\frac{M_l^{\rm obs}m_{n,l}^{\rm obs}-\sum_k\gamma_kM_l^km_{n,l}^{k}}{\Delta
(M_l^{\rm obs}m_{n,l}^{\rm obs})}\Bigg)^2
\end{equation}
under the constraints that for each $k$ and each $l$:
\begin{equation}
\label{eq:weights}
\left\{
\begin{array}{l}
|M_l^{\rm obs}-\sum_k\gamma_kM_l^k|\leq\Delta M_l^{\rm obs}\\
\gamma_k\ge 0
\end{array} \right.
\end{equation}
where $M_l^k$, $M_l^{\rm obs}$ are the fractions of the projected mass of the tracer contained within $l$th bin for
$k$th orbit or from the observations and $m_{n,l}^k$, $m_{n,l}^{\rm obs}$ are $n$th proper moments. $\Delta$ denotes
the measurement uncertainty associated with a given parameter. The velocity moments are weighted with the projected
masses and to derive the errors we treat both quantities as independent.

We assume that tracer particles are massless and orbit in the potential of their dark matter halo. However, if the
mass-to-light ratio of stars is constant with radius, then $M_l^{\rm obs}\equiv N_l^{\rm obs}$, where $N_l^{\rm obs}$ is
a fraction of tracer particles contained within the $l$th bin. Therefore in this study by `projected mass' we mean
$N_l^{\rm obs}$.

As it has been already pointed out by \citet{breddels_2013}, the 4th velocity moment is not independent of the 2nd
moment. Therefore, the kurtosis ($\kappa=m_4/m_2^2$), which is not correlated with the 2nd moment,
is preferred as a kinematical parameter in dynamical studies. However, the kurtosis cannot be used
as a constraint for the Schwarzschild modelling, as it would not be linear in the orbital weights. Consequently, we 
proceed neglecting the possible correlations and using the proper 4th moment $m_4$.

We used an additional constraint on the sum of the weights. As the weights have the physical meaning of the amount of
mass assigned to the corresponding orbits, they should sum up to unity as long as an orbit library covers
the whole (or typically in numerical studies $\geq 99.9\%$, which holds for our library) deprojected mass of the tracer:
\begin{equation}
 \sum_k\gamma_k=1.
\end{equation}

The minimization of the objective function under the equality and inequality constraints has been executed
using quadratic programming as
implemented in the CGAL\footnote{\url{www.cgal.org}} library
(The Computational Geometry Algorithms Library, \citealt{cgal:eb-15b}).

We calculate the resulting anisotropy $\beta$ in the $l$th bin by assuming that:
\begin{equation}
 \beta_l=1-\frac{\sum_k\gamma_kM_{{3D}, l}^k(\sigma_{\theta, l}^k)^2+\sum_k\gamma_kM_{{3D}, l}^k(\sigma_{\phi,
l}^k)^2}{2\sum_k\gamma_kM_{{3D}, l}^k(\sigma_{r, l}^k)^2}
\end{equation}
where $\sigma^k_{(r,\,\theta,\,\phi),l}$ are the components of the velocity dispersion in the spherical coordinate
system for the $k$th orbit calculated in the $l$th spatial bin.

\section{Large data samples}
\label{large_sample}

In this section we present the application of our method to the data derived from all available 
stellar particles (see section \ref{tracer}) contained within a cylinder of radius $R=6$\,kpc, corresponding to 
(2.11--2.24)$\times 10^5$ particles, depending on the model.

\subsection{Known mass profile}

First, we check how reliably we can recover the anisotropy profile alone. For this purpose we assumed that the correct
mass profile (see section \ref{data}) was known and performed the fitting of the mock data to the orbit library
calculated in the potential generated by this distribution. In each of the 30 radial bins we applied 6 constraints 
following eq.\,(\ref{eq:weights}) and (\ref{eq:fit}): the
rigid lower and upper boundaries for the mass projected along the line of sight and the $\chi^2$ fit for the velocity
moments 1-4 given with eq.\,(\ref{eq:mom}) and (\ref{eq:mean}). We assumed Poissonian errors for the projected masses
and the theoretical sampling errors (\citealt{kendall_1977}, \citealt{harding_2014}) of standard deviation:
\begin{equation}
 \Delta \sigma =\frac{\sigma}{\sqrt{2(n-1)}}
\end{equation}
where $n$ is the size of a sample, skewness:
\begin{equation}
 \Delta \gamma =\sqrt{\frac{6n(n-1)}{(n-2)(n+1)(n+3)}}
\end{equation}
where $\gamma =m_3/m_2^{3/2}$ and kurtosis:
\begin{equation}
 \Delta \kappa =2\Delta \gamma \sqrt{\frac{n^2-1}{(n-3)(n+5)}} 
\end{equation}
propagated back to the velocity moments $m_2$, $m_3$ and $m_4$. The formulae are derived under the assumption that the parent 
distribution is normal. {\it A priori} it may not be true for our models. However, as the data samples we use are large, the 
crude estimates of sampling errors are sufficient in order to test the method. We present a precise study of sampling errors 
and their application in section \ref{small_sample}.

By definition, $m_1(R)\equiv 0$ up to numerical precision (typically $10^{-14}-10^{-17}$) so we applied a fixed value 
of $\Delta m_1=0.001$ everywhere.

\begin{figure}
\includegraphics[trim=50 40 110 55, clip, width=\columnwidth]{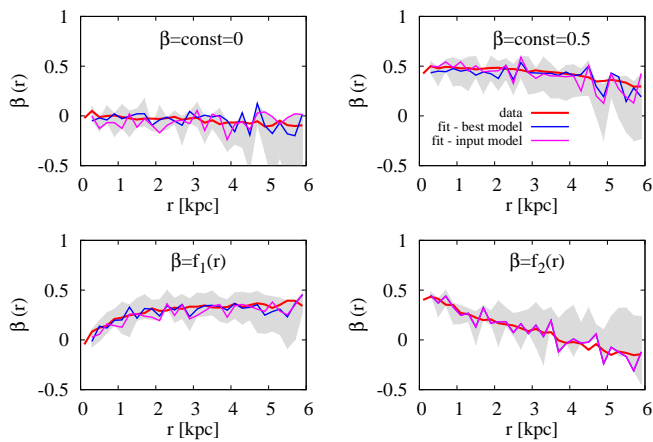}
\caption{The anisotropy parameter profiles for all particles from the simulations for the four models. In red
we present the values of direct measurements from the full data and in magenta and blue the fits obtained with
the assumption of the correct mass profile and for the best-fitting mass profile, respectively. The shaded regions
correspond to the extreme values for the mass profiles recovered within the $1\sigma$ confidence level.}
\label{fig:beta_all}
\end{figure}

The results are presented with magenta curves in Fig.\,\ref{fig:beta_all} together with the true values (in red)
calculated from the full 6D information about the particles. The four panels refer to the models with different
anisotropy. The accuracy of the recovered profiles is remarkable, with only some noise.

We also need to comment on the missing results for the innermost bins in Fig.\,\ref{fig:beta_all} and the following
figures presenting recovered anisotropies. In our opinion the innermost bin is underconstrained, having only one
neighbour and many orbits contained entirely within one or two bins (which is not the case for the outermost bin) so
one cannot rely on the results in this bin. This effect manifests itself notably for the small data samples leading to
enormous (when compared to other bins) and highly non-Gaussian errors. A problem of the innermost bin has been already
reported by \citet{breddels_2013}, however with a different justification.

\subsection{Unknown mass profile}
\label{profiles_30}

We also examined the possibility of recovering the underlying mass profile for the four models. We constructed a
grid of profiles given with eq.\,(\ref{eq:mass_profile}) with the values of the virial mass and
concentration spaced linearly in the ranges: $M_v\in[0.2,\,3]\times 10^9\mathrm{M}_{\sun}$ and $c\in[8,\,27]$. The
parameters of cut-off in the mass profile, i.e. $N$ and $r_c$, were in each case adopted so that the profile and its
first derivative at the distance of the virial radius were continuous. For each profile we integrated the library of
orbits as explained in section \ref{orbits}.

\begin{figure}
\includegraphics[trim=0 0 0 120, clip, width=\columnwidth]{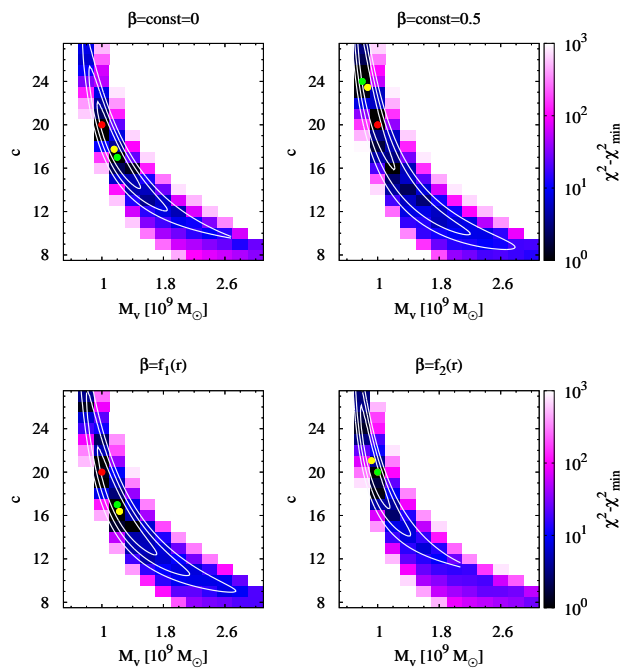}
\caption{The maps of the $\chi^2$ values relative to the minima of fitted surfaces for four anisotropy models on the grids
of different mass profiles. The global minima are marked with yellow dots and the true values with red (in the case of
the decreasing anisotropy they overlap). Green points
indicate the best-fitting mass profiles, i.e. the profiles on the grid closest to the global minima along the contours of
equal $\Delta\chi^2$ plotted with white curves.}
\label{fig:chi_all}
\end{figure}

Similarly to the procedure described in the previous section, for each model we used the projected mass and the
velocity moments 1-4 in 30 spatial bins as constraints to fit the orbit library. The absolute values of the $\chi^2$
function, eq.\,(\ref{eq:fit}), have been saved and we compare them in Fig.\,\ref{fig:chi_all}, where each panel
corresponds to a different anisotropy model. The logarithmic colour scale represents the differences of the values of
$\chi^2$ relative to the minimum ($\chi^2_{\rm min}$) of the fitted two dimensional surfaces of the 4th order ($\propto
M_v^2c^2$). The minima are marked with yellow dots. The white lines indicate the contours of equal $\Delta
\chi^2=2.3,\,6.17,\,\mathrm{and}\,11.8$ corresponding to $1,\,2,\,\mathrm{and}\,3\sigma$ confidence levels for two
degrees of freedom \citep{NR} also based on the fitted surfaces. The concentration is constrained much more poorly than
the virial mass as it is a very sensitive parameter.

We have found the fitting of a surface necessary in order to derive a global minimum and contours of equal
$\Delta\chi^2$ as the Schwarzschild method is severely influenced by numerical effects and therefore one should
consider trends rather than particular values. The discretization appears on many levels and cannot be avoided: a smooth
distribution function of a galaxy is represented by a finite set of deltas (orbits), continuous orbits are represented
by a finite set of timesteps and the spherical symmetry is represented by a finite set of rotations of the orbits.

The true values of the mass profiles are marked with red dots and in each case lie within the $1\sigma$ regions.
In order to avoid calculating libraries for the minima of the fitted surfaces which do not correspond 
to any models on the adopted mass grid, we identified the models on the grid closest to the global minima along the 
contours of equal $\Delta\chi^2$ as the best-fitting models and marked them in Fig.\,\ref{fig:chi_all} with green dots. 
For the galaxy with the decreasing anisotropy profile the best-fitting and true density models overlap. The resulting 
anisotropy profiles for the best-fitting models are plotted in Fig.\,\ref{fig:beta_all} in blue.

Using the $1\sigma$ regions of the recovered mass profiles we estimated errors on the values of the recovered
anisotropy by taking in each bin the extreme values among the mass profiles within $\Delta \chi^2=2.3$. The results are
shown in Fig.\,\ref{fig:beta_all} as the shaded regions, following well the values and shapes of the true anisotropies.

Despite the mass-anisotropy degeneracy, with our method we have recovered the proper values of both the anisotropy
and the mass profile: $M_v = 10^9\mathrm{M}_{\sun}$ and $c = 20$ for each model with high accuracy, independently of
the underlying anisotropy.

The anisotropy can be equally well recovered using only the 2nd and 4th velocity moments. However, the addition of
the 1st and 3rd moments to the fit, which is not typical, influenced the recovery of the mass profile, providing
an estimate with
a higher confidence. As it will prove to be of great importance in the next section, we decided to use all the
moments.

\section{Small data samples}
\label{small_sample}

Section \ref{large_sample} shows the strength of the Schwarzschild modelling in breaking the mass-anisotropy
degeneracy. The weak point of this argument is the amount of data we used, which is impossible to achieve in the case of
dwarf galaxies of the LG. In order to test our method on realistic samples, for each of the
galaxy models we randomly chose 100\,000 stellar particles with positions (in observations those are stars with only
photometry measured) and 2\,500 particles with positions and line-of-sight velocity (corresponding to spectroscopic
data) contained within the projected radius of 6\,kpc and binned them in 10 radial bins spaced linearly.

It is typical for the treatment of small data samples to adjust binning to the data by fixing the number of stars
in the bin. However, binning is then based on a particular data sample, not the 
parent distribution. Therefore, when fixing a number of particles we impose sampling errors on borders of the adopted binning, 
as different random samples would result in different spatial partitions. 
Since this effect is impossible to correct for in our method, we decided to keep the predefined binning fixed in radius.

\subsection{Sampling errors}
\label{simulations}

For observational data the sampling errors dominate over the measurement errors of line-of-sight velocities for single
stars and are therefore the main source of uncertainties imposed on the velocity moments. In order to estimate them
properly we ran Monte Carlo (MC) simulations deriving the sampling errors for various parameters ($\beta$, $m_2$,
$m_3$, $m_4$) for the adopted spatial binning and for each halo model as a function of a number of particles in each
bin in the range of sample sizes from $40$ to $550$ particles with the step of $30$. Between the nodes of such a dense
grid the errors can be interpolated linearly. This allowed us to apply our method regardless of the size of the current
data sample.

\begin{figure}
\includegraphics[trim=50 40 110 55, clip, width=\columnwidth]{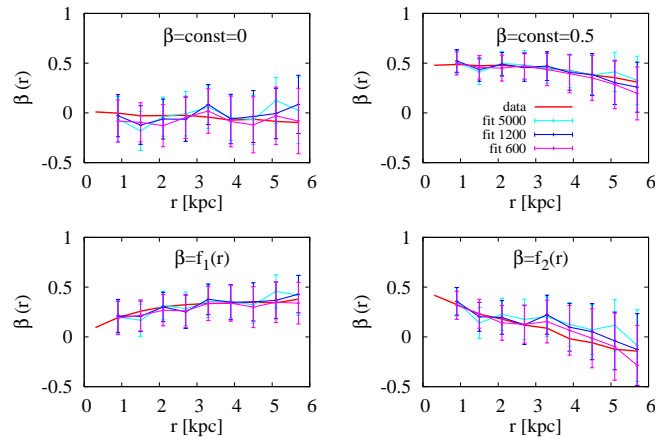}
\caption{The mean values with $1\sigma$ error bars resulting from the MC simulations. In red we present the values
based on direct measurements from the full data and in cyan, blue and magenta results for the fits done with libraries of
5\,000, 1\,200 and 600 orbits, respectively.}
\label{fig:beta_errors}
\end{figure}

We took advantage of it when running the next set of tests. For each halo we randomly selected our data samples,
assigned the errors and fitted the orbit library, repeating the procedure 10\,000 times. The resulting distributions
of the anisotropy profiles were fitted with Gaussians in order to derive the mean values and the $1\sigma$ deviations,
in each spatial bin separately. The results for all models are shown in Fig.\,\ref{fig:beta_errors} in cyan, dark
blue and magenta, depending on the number of orbits used for the fit. We used: 5\,000 (a library larger than the
one used in the study; 200 values of energy $\times$ 25 values of angular momentum), 1200 (the default library;
100 $\times$ 12) and 600 (a smaller one; 75 $\times$ 8), respectively. In red we present
the true values for the anisotropy derived from full 6D information about the particles.

We note the rather high accuracy of the obtained mean values and the relatively small errors which allow to clearly
differentiate between different models of anisotropy. The uncertainties in the resulting anisotropy, derived with our
method from the projected positions and line-of-sight velocities are only $\sim2\times$ larger than the sampling errors
for the anisotropy calculated from full deprojected positions and three-dimensional velocity vectors, i.e. variations
in the anisotropies originating from taking random small samples of particles (compare with Fig.\,\ref{fig:beta_ss}).
The growth of the errors with radius is a consequence of the decreasing number of particles in subsequent bins. The
offset between the true values and the mean ones is a consequence of the changes of anisotropy outside the modelled
area. Also the mean values seem to follow a weak trend in which the anisotropy is recovered more 
precisely for fewer degrees of freedom (less orbits in a library), whereas in all cases the deviations remain roughly 
the same. We may trace it back to the fitting procedure and the applied rigid constraints which diminish the impact of 
the kinematical constraints.

\subsection{Examples of data modelling}

\begin{figure}
\includegraphics[width=\columnwidth]{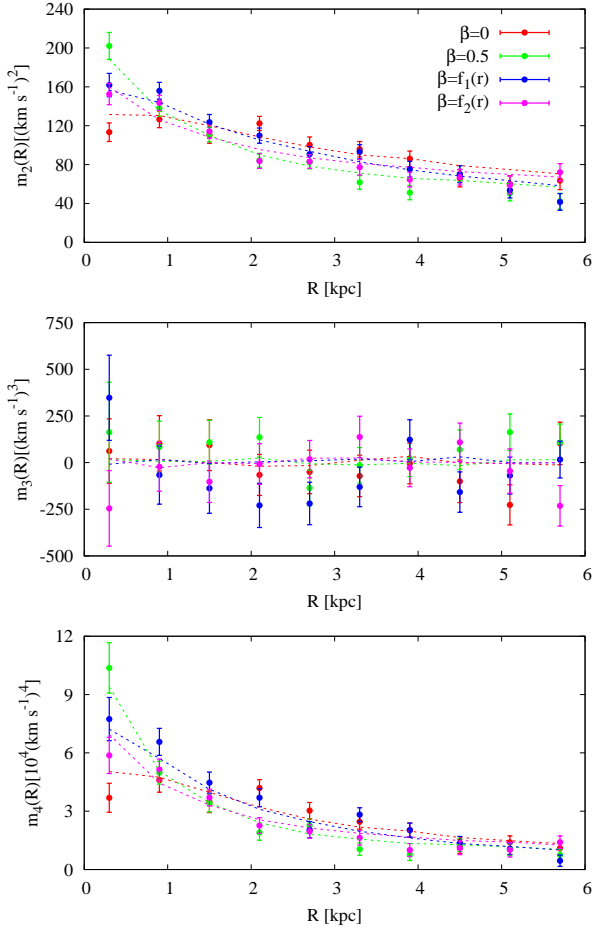}
\caption{The values of the 2nd, 3rd, and 4th velocity moments (top to bottom panels, respectively) for the four models:
$\beta=0$ (red), $\beta=0.5$ (green), increasing $\beta$ (blue) and decreasing $\beta$ (magenta). The points with the
$1\sigma$ error bars represent the values for the random samples of $2\,500$ stars, while the thin dashed lines show
the results based on all stellar particles from the simulations with the same binning.}
\label{fig:mom_ss}
\end{figure}

In this section we present the results of modelling four sets of mock data for 100\,000 (positions) and 2\,500
(velocities) randomly selected particles, one for each model. Fig. \ref{fig:mom_ss} shows the kinematics of the samples
as points with $1\sigma$ errors derived in section \ref{simulations} and for comparison the same parameters for all
stellar particles from the simulations with the same binning as thin dashed lines. Colours denote different anisotropy
models: red for the isotropic case $\beta=0$, green for $\beta=0.5$, blue for the growing anisotropy and magenta for
the decreasing one.

\begin{figure}
\includegraphics[trim=50 40 110 0, clip, width=\columnwidth]{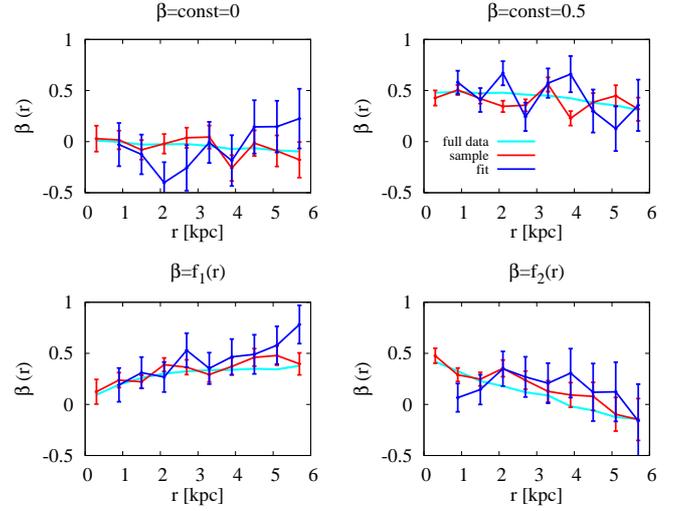}
\caption{The anisotropy parameter profiles for random samples of particles for the four models. In cyan we present the
values based on direct measurements from the full data from the simulations, in red the values from the
used samples and in blue the fits with the assumption of the correct mass profile. The error bars denote the
$1\sigma$ errors.}
\label{fig:beta_ss}
\end{figure}

The profiles of the anisotropy are shown in Fig. \ref{fig:beta_ss} where the red points correspond to true values from
the data sample and blue ones to the recovered anisotropy, both with the errors calculated in section \ref{simulations}.
The smooth cyan curves present the values for all stellar particles from the simulations for comparison.

\subsection{Recovering the mass profile}

\begin{figure}
\includegraphics[trim=0 0 0 120, clip, width=\columnwidth]{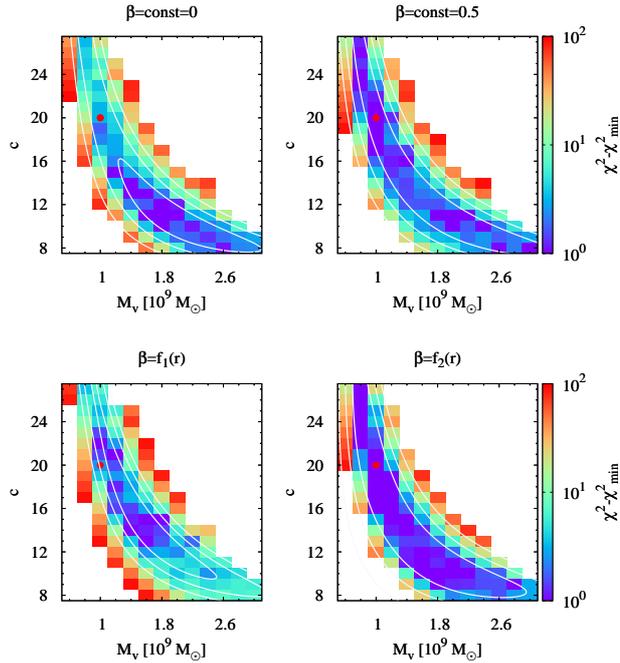}
\caption{Maps of the $\chi^2$ values relative to the minima of the fitted surfaces for four 
models on the grid of
different mass profiles for the small data samples of 100\,000/2\,500 particles. Thin white lines indicate the
contours of equal $\Delta \chi^2$ corresponding to $1,\,2,\,3\sigma$ confidence levels. The true values of the
density profiles are marked with red dots.}
\label{fig:chi_ss}
\end{figure}

As for the large data samples, we also studied the reliability of recovering the mass profile for our small data
samples, fitting libraries of 1200 orbits integrated in the potentials generated by the mass profiles described in
section \ref{profiles_30}. We present the resulting colour maps of $\Delta \chi^2=\chi^2-\chi^2_{\rm min}$ as a
function of virial mass and concentration in Fig.\,\ref{fig:chi_ss}. As it has been done in section \ref{profiles_30},
we derived the minima and $1,\,2,\,\mathrm{and}\,3\sigma$ confidence levels by fitting two-dimensional surfaces to the
$\chi^2$ maps. For the stable models ($\beta=0$ and $\beta=f_1(r)$, see section \ref{tracer}) the virial masses are
overestimated whereas the concentrations are underestimated, covering the true profiles within $2\sigma$. This suggests
that the particles `feel' a slightly different potential.

For $\beta=0$ we have confirmed that this is not an outlier, i.e. an unfortunate random sample, 
in a test in which we studied 10\,000 different random samples (as for the sampling errors of recovered anisotropy,
section \ref{simulations}), fitting all the orbit libraries and calculating the mean $\chi^2$ values for each library.
This behaviour might be a consequence of our choice of the outer radius of the data sets as the anisotropy grows
rapidly outside it (see Fig.\,\ref{fig:beta_init_fin}). The particles which are in large physical distance from the
centre still enter our calculations since their projected distances are smaller so that the small samples are
contaminated by the particles on radial orbits. Higher values of line-of-sight velocities cause the line-of-sight
velocity dispersion to grow, which (under the assumption of the dynamical equilibrium) leads to overestimation of the
total mass. Unfortunately, those particles also affect the values of sampling errors, enlarging them. It was not the
case for the large samples as `contaminants' were outnumbered by `well-behaved' particles and the errors were
calculated analytically.
For the unstable models the situations is even worse. Larger sampling errors result in poorly constrained density profile,
spanning large area of our grid.

\begin{figure}
\includegraphics[trim=50 40 110 0, clip, width=\columnwidth]{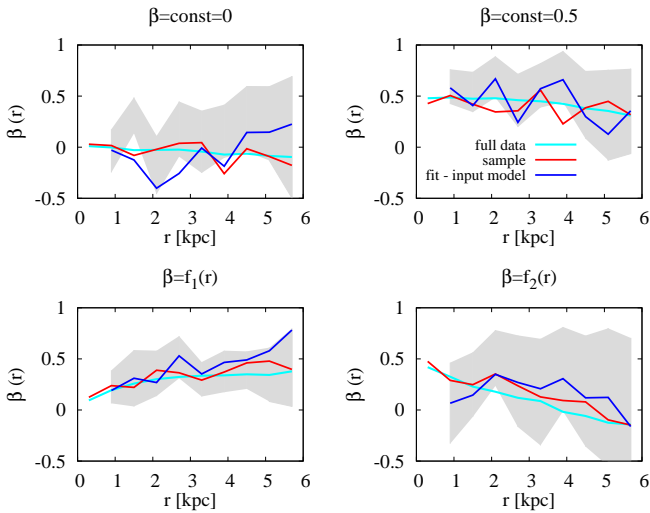}
\caption{The anisotropy parameter profiles for random samples of particles for the four models. The shaded regions
correspond to the extreme values for the mass profiles recovered within the $1\sigma$ confidence level. In cyan we
present the values based on direct measurements from the full data from the simulations, in red the values
based on the used samples and in blue the values for the true mass profile.}
\label{fig:beta_shade}
\end{figure}

As the $1\sigma$ regions are large and may not be very accurate, we decided not to identify the minima as the
best-fitting models. Instead, in Fig.\,\ref{fig:beta_shade} we present only the ranges (shaded regions) of the
values of anisotropy spanned by the results for the density profiles within $1\sigma$. The true values of anisotropy are
presented with cyan and red lines for all particles and small samples, respectively. The blue lines correspond to the
results for the true mass profiles for comparison.

Despite the wide ranges of the similarly plausible density profiles,
the derived anisotropy intervals are not much larger than the sampling errors for the known mass distribution.
On average the deviations are larger by $35\%$ for the stable models (with $\beta=0$ and
with the growing profile), $70\%$ for $\beta=0.5$ and $132\%$ for the model with decreasing anisotropy. Nevertheless
the intervals include the correct values and follow the general behaviour of the anisotropy profiles. Our approach
is simplistic and does not provide a full picture
as the recovered anisotropy for each mass profile is additionally subject to the sampling errors as presented in
section\,\ref{simulations}.

\section{Summary}
\label{summary}

We have presented a study aimed at determining the efficiency of recovering the anisotropy and density
profiles by the application of the Schwarzschild modelling method to a set of four dwarf galaxies obtained
from the numerical realizations of NFW dark matter haloes by marking particles described with a S\'ersic profile
and following their evolution in isolated haloes in order to achieve equilibrium. The models shared
the same spherically symmetric density profile but differed in the orbit anisotropy, covering a wide class of possible
profiles and therefore allowing for a thorough test of the scheme. We have tested in total four models of anisotropy,
two constant with radius $\beta=0$ (isotropic model) and $\beta=0.5$, and two with anisotropy varying with
radius, one growing and one decreasing.

We performed our tests applying two different approaches, and in addition using two types of samples,
which we called \textit{large} and \textit{small}. The large sample contained over 211\,000 
particles within the projected radius of $6$\,kpc used as the outer boundary of the mock data. As the
small samples we used subsamples of the large ones by randomly choosing 100\,000 particles with
positions and 2\,500 particles with positions and line-of-sight velocities imitating the best data samples
currently available for dwarf galaxies of the LG.

First, we assumed we knew the density profile exactly and performed the fitting of the observables in order to
retrieve only the anisotropy profile. Our results for the large samples show that the anisotropy
can be recovered with very high accuracy independently of its profile. We have demonstrated that also for the small
samples our method provides interesting results. We carried out a set of MC simulations in order to determine the
sampling errors imposed on the recovered values of the anisotropy, deriving the mean value over the profiles and the
radial bins of $\sigma_{\beta}=0.2$, only $\sim2\times$ larger than the mean sampling error for the anisotropy. Such small
errors enable us to clearly distinguish between the different models
of anisotropy we used, proving the strength of the Schwarzschild method in this respect.

In the second approach we tested how precisely we can recover both the mass and anisotropy profiles. We assumed that
the profile was given by the NFW formula with a cut-off at the distance of the virial radius and we constructed a
grid of orbit libraries by varying the virial mass and the concentration. For the large data samples we have recovered
the true mass profile for each halo model within the confidence level of $1\sigma$ whereas for the small samples the
parameters of the density profiles were strongly degenerated, resulting in extended regions of possible values.
However, the correct values were included within at least $2\sigma$ confidence regions.

Finally we have calculated the uncertainties associated with the anisotropy and coming from the uncertainty of the
recovered mass profile. They are not much larger (by only $35\%$ for the two reliable models) than the sampling errors
derived from our MC simulations demonstrating that the unknown mass distribution affects anisotropy similarly to the
limited amount of data.

\section{Discussion}
\label{discussion}

The attempts to recover the anisotropy for dwarf galaxies with Schwarzschild modelling were already made for
Fornax dSph \citep{jardel_2012} and Draco dSph \citep{jardel_2013} but without a clear demonstration that the
undertaken procedures actually work. Here, we have filled this gap by showing that the anisotropy, regardless of its
profile, can indeed be recovered by this method.

For the purpose of simplicity we have tested the method on numerical realizations of dark matter haloes only,
therefore neglecting the stellar component. However, as dwarf galaxies are believed to be highly dark matter dominated,
at first approximation we may assume that the influence of stars on the dynamics of the system is in fact
negligible and stars move in the potential generated by the distribution of dark matter. However, we need to bear
in mind that as a result we obtain the \textit{total} mass profile in which the orbit library has been integrated
and the anisotropy profile of the \textit{tracer}. We plan to implement the stellar mass to the fit in future extensions
of our models by quantifying total mass in terms of the mass-to-light ratio varying with radius $\Upsilon(r)$.

Yet another complication may arise from the stellar mass-to-light ratio $\Upsilon_\star$ varying with radius. However,
as there is no strong evidence for its gradients in dwarfs (even for the ones with multiple stellar
populations) it is typical to assume that the parameter is constant. It has been derived for many galaxies of the LG
\citep{mateo_1998}.

\citet{breddels_2013} did similar work to the one presented here, testing the Schwarzschild modelling on the mock
Sculptor dSph with anisotropy assumed to be constant with radius, $\beta=-1$. These authors obtained good estimates
of the mass profile, which we were not able to reach for our haloes, so we conclude that the parametrization of the mass
profile may play a role in recovering precise values as the stars, concentrated at the centre of the dark matter halo,
do not feel the mass distribution at the virial radius. After reparametrization of the density profiles, we found that
the mass contained within 6\,kpc ($M_{6{\rm kpc}}$) was constrained much better (overestimated by no more than 50\%)
but the characteristic radius of the NFW profile ($r_s$) could be overestimated as much as 4 times for the unstable
models and small data samples. Also the underlying anisotropy profile (tangential vs.
our radial) may affect the quality of the mass profile recovery.

As the final remark we would like to comment on two tools often applied to the orbit superposition method in
order to enforce smoother, more physical distribution function. The first tool is the regularization
(\citealt{vdMarel_1998}, \citealt{valluri_2004}), which imposes a penalty term restraining the values of the weights
of the consecutive orbits (in the energy, angular momentum or both) so that they do not differ \textit{too much}. The
expression `too much' is not precise and the strength of the regularization is a moot point. The regularization worsens
the quality of the fit and for example \citet{rix_1997} imposed a constraint on the resulting $\chi^2$ value such that
it was not to be changed by the regularization by more than $\Delta \chi^2<1$ when compared with the non-regularized
case. \citet{breddels_2013} used an arbitrary value of the regularization strength which according to them
worked well. In our opinion such approaches do not affect the distribution function of the system sufficiently to
classify it as smooth while much higher values of regularization strength result in diminishing the role of the
observational constraints and cause the regularization to dominate. Therefore, following the example of
\citet{yildirim_2015}, we have decided not to apply the regularization at all.

The other tool is the \textit{dithering} (\citealt{rix_1997}, \citealt{breddels_2013}) in which one orbit in a library
is constructed as a compound of a few (typically 5-8) orbits with close values of energy and angular momentum. 
It might be treated as a kind of strong regularization on the subgrid level, since `suborbits' 
are assigned equal weights, but it is much more quantitative and we do not discard the possibility of including 
it in our modelling in the future.

\section*{Acknowledgements}

This research was supported in part by the
Polish Ministry of Science and Higher Education under grant 0149/DIA/2013/42 within the Diamond Grant Programme for
years 2013-2017 and by the Polish National Science Centre under grant 2013/10/A/ST9/00023. We thank R. Wojtak for
sharing $N$-body realizations of dark matter haloes with different anisotropy profiles. MV acknowledges support
from HST-AR-13890.001, NSF awards AST-0908346, AST-1515001, NASA-ATP award NNX15AK79G.

\bsp
\label{lastpage}
\end{document}